# Anisotropic Structural and Magnetic Response in $Sm_2Ir_8Al_{31}$: Al-Sublattice Reconstruction upon Sm Substitution

Mingyu Xu[1], Weiwei Xie[1*]

1. Department of Chemistry, Michigan State University, East Lansing, MI, 48824 USA

[*]Corresponding author: Dr. Weiwei Xie (xieweiwe@msu.edu)

***Abstract***

Chemical pressure plays a critical role in determining the crystal structures and physical properties of intermetallic compounds. Here, we report the synthesis, crystal structure, and physical properties of $Sm_2Ir_8Al_{31}$, a structural analog of $UIr_4Al_{15}$, in which the smaller $5f$ uranium atoms are replaced by larger $4f$ samarium atoms. Single-crystal X-ray diffraction reveals a tetragonal $\sqrt{2} \times \sqrt{2} \times 1$ superstructure with space group $P\bar{4}2_1m$, derived from a $UIr_4Al_{15}$-like average structure with space group $P4_2/nmc$. Comparison with $UIr_4Al_{15}$ shows that chemical pressure from U-to-Sm substitution produces a strongly anisotropic structural response rather than a uniform expansion of the Ir-Al framework. While one Ir coordination environment undergoes only modest changes, the other exhibits substantial local rearrangement. The Al sublattice primarily accommodates these structural distortions, with selected Al sites showing positional splitting, anisotropic atomic displacement ellipsoids, and partial occupancies, resulting in an Al-rich composition of approximately $Sm_2Ir_8Al_{31}$. Magnetic susceptibility measurements reveal a direction dependence, including a broad anomaly near 50 K for selected magnetic-field orientations. However, the absence of corresponding anomalies in heat capacity, electrical resistivity, and lattice parameters suggests that this feature is more consistent with an anisotropic magnetic crossover than a bulk thermodynamic phase transition. These results highlight how chemical substitution at the *f*-element site can induce Al-sublattice reconstruction and modify the magnetic response within a closely related Ir-Al structural framework.

## Introduction

Intermetallic compounds containing rare-earth or actinide elements and transition metals provide a rich platform for investigating the interplay among crystal structure, chemical bonding, and electronic and magnetic properties, particularly the interactions between *f*- and *d*-electron states. In these materials, the *f*-element's atomic size and electronic configuration, together with the transition metals' bonding characteristics, can strongly influence the structural stability, local coordination environments, and electronic properties of the surrounding framework[1–6]. A representative example is the $RT_2Al_{20}$ family (R = rare-earth element; $T$ = transition metal), such as $PrV_2Al_{20}$, in which rare-earth atoms occupy highly symmetric, oversized Al cages with 16-fold coordination[7–14]. The relatively weak confinement of the rare-earth ions within these cages can give rise to low-energy vibrational dynamics[15], while the interplay among crystal-field effects, conduction electrons, and localized *f* electrons produces a variety of strongly correlated electronic phenomena, including heavy-fermion behavior, multipolar ordering, and quantum criticality[16–29]. These examples highlight how local atomic environments and chemical bonding can govern the emergence of complex electronic and magnetic states.

$UIr_4Al_{15}$ crystallizes in a tetragonal structure consisting of an extended Ir-Al framework, with U atoms occupying well-defined crystallographic sites[30,31]. Uranium's relatively extended 5*f* orbitals can hybridize with surrounding ligand states, making its local coordination environment and electronic properties sensitive to changes in chemical bonding. Substituting U with a rare-earth element provides an opportunity to investigate the combined effects of chemical pressure and *f*-electron character on the structural stability and physical properties of the Ir-Al framework. Replacing U with Sm introduces a larger atomic species with more localized 4*f* electrons and potentially strong magnetic anisotropy associated with crystal-field effects[32]. Such substitution can modify both the framework's geometry and electronic interactions, potentially leading to local structural distortions and distinct magnetic responses.

Here, we report the synthesis, crystal structure, and physical properties of $Sm_2Ir_8Al_{31}$, a structural analog of $UIr_4Al_{15}$ that preserves the underlying Ir-Al framework while substantially modifying the Al sublattice. Single-crystal X-ray diffraction reveals a tetragonal structure in space group $P\bar{4}2_1m$. A detailed structural comparison shows that replacing U with Sm does not simply produce a uniform contraction or expansion of the framework. Instead, the Ir-Al network shows a highly anisotropic, site-dependent structural response. While one Ir coordination environment remains relatively unchanged, the other undergoes substantially larger local rearrangements. The Al sublattice primarily accommodates these distortions, with several Al sites showing positional splitting, pronounced anisotropic displacements, and partial occupancies. Consequently, the resulting composition, $Sm_2Ir_8Al_{31}$, contains more Al than expected from direct stoichiometric substitution in $UIr_4Al_{15}$. The physical properties of $Sm_2Ir_8Al_{31}$ further reveal distinct responses to replacing the actinide element with a rare-earth element. Unlike $UIr_4Al_{15}$, which exhibits temperature-dependent anomalies in its lattice parameters and physical properties[30], $Sm_2Ir_8Al_{31}$ shows no comparable features over the investigated temperature range. Its heat capacity can be described using a two-Debye model, with no clear evidence of a bulk thermodynamic phase transition. In contrast, magnetic susceptibility measurements reveal pronounced magnetic anisotropy, including a broad, orientation-dependent feature near 50 K. The absence of a corresponding heat-capacity anomaly suggests that this behavior is more consistent with an anisotropic magnetic crossover, potentially associated with crystal-field effects or short-range magnetic correlations[32–34], rather than conventional long-range magnetic ordering. This study highlights the interplay among chemical pressure, local chemical bonding, and *f*-electron character in determining the structural stability and anisotropic physical properties of complex rare-earth and actinide intermetallic compounds.

## Experimental and Computational Details

**Crystal Synthesis:** Single crystals of $Sm_2Ir_8Al_{31}$ were grown using a high-temperature Al self-flux method[35]. Sm pieces (Thermal Scientific, pieces,99.9% REO), Ir powder (Mennica Metale, 99.9%), and Al shot (Alfa Aesar, 99.9% metal basis) with an atomic ratio of 1:1:50 were loaded into an $Al_2O_3$ crucible set[36] and sealed in an evacuated quartz tube. The quartz tube was heated to 1180 °C, held to ensure complete homogenization, and then slowly cooled down to 720 °C over 86 h. At 720 °C, centrifugation removed the remaining molten Al flux, yielding well-formed single crystals of $SmIr_4Al_{15}$. Typical crystals have dimensions of approximately 0.2-0.5 mm, as shown in the inset of **Figure 2**, and are stable in air under ambient conditions.

**Single-Crystal X-ray diffraction (SCXRD) and Powder X-ray diffraction (PXRD):** A $Sm_2Ir_8Al_{31}$ single crystal (0.15 × 0.12 × 0.05 $mm^3$) was mounted on a nylon loop with Paratone oil and measured at room temperature; then the same crystal was mounted on a nylon loop with nail polish and measured from 100 K to 350 K using a Rigaku XtaLAB Synergy Dualflex diffractometer with a HyPix detector. Data were collected using ω scans with Mo Kα radiation (λ = 0.71073 Å) from a micro-focus sealed tube (50 kV, 1 mA). Data collection strategies were optimized using CrysAlisPro (v1.171.43.144a). Data reduction included Lorentz and polarization corrections, numerical absorption correction based on Gaussian integration over a multifaceted crystal model[37], and empirical spherical harmonics correction using the SCALE3 ABSPACK algorithm[38]. Structures were solved and refined using Olex2 with the SHELXTL package[39–41]. Refinement details and atomic coordination parameters are summarized in **Tables S1** and **S2**, respectively**.** To examine the phase purity information, the powder X-ray diffraction (PXRD) analysis was performed. The crystals were ground using an agate mortar and pestle to achieve a homogeneous powder. The PXRD data acquisition spanned a $2\theta$ range from 15° to 85°, utilizing incremental steps of 0.01° and a fixed dwell time of 3 seconds per step. These measurements were conducted using a Rigaku MiniFlex II powder diffractometer, employing Bragg-Brentano geometry coupled with Cu Kα radiation (λ = 1.5406 Å). The refinement of the powder X-ray data was executed using the GSAS-II software suite[42].

**Scanning Electron Microscope and Energy-Dispersive Spectroscopy:** The phase composition was analyzed employing a JEOL 6610LV scanning electron microscope equipped with a tungsten hairpin emitter (JEOL Ltd., Tokyo, Japan). For elemental analysis, energy-dispersive X-ray spectroscopy (EDS) was conducted utilizing an Oxford Instruments AZtec system (Oxford Instruments, High Wycomb, Buckinghamshire, England), operating software version 3.1. This setup included a 20 mm$^2$ Silicon Drift Detector (SDD) and an ultra-thin window integrated with the JEOL 6610LV SEM. The single crystals of $SmIr_4Al_{15.4}$ were attached to carbon adhesive tape and introduced into the SEM chamber for examination at an accelerating voltage of 20 kV. Data acquisition entailed collecting spectra at multiple points along the individual crystals over an optimized timeframe. Quantitative compositional analysis was performed using SEM Quant software, which applies matrix-effect corrections to the intensity measurements.

**Magnetization, Electrical Transport and Heat Capacity Measurements:** Magnetization, electrical resistivity, and specific heat measurements were performed using Quantum Design (QD), Magnetic Property Measurement Systems (MPMS3), and Physical Property Measurement Systems (PPMS). During magnetization measurements of single crystals, the field was applied along the a-axis, as confirmed by single-crystal X-ray diffraction. The samples are placed between two collapsed plastic straws, with a third uncollapsed straw providing support as an outer sheath or a quartz sample holder. We performed some angle-dependent magnetic measurements using a cut straw at predefined angles. We fixed the samples to the angled cross-section of the straw, while a second straw provided mechanical support. Samples were fixed on the straw with GE-7031 varnish. We measured electrical resistivity using a standard four-probe configuration. We measured heat capacity using the relaxation technique over the PPMS's available temperature range.

## Results and Discussion

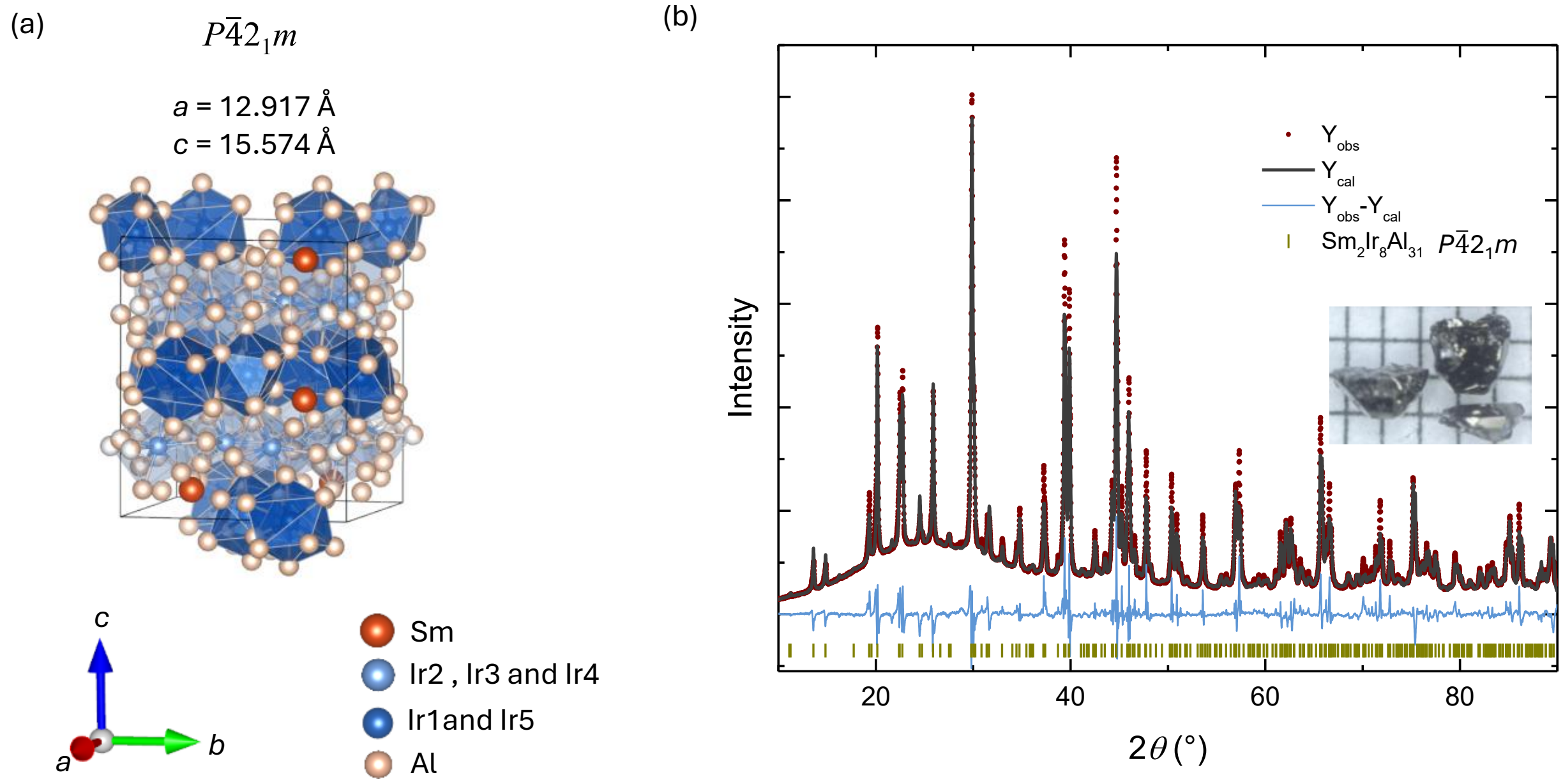

**Fig. 1| Crystalline structure of $Sm_2Ir_8Al_{31}$.** (***a***) Crystal structure of $Sm_2Ir_8Al_{31}$. (***b***) Powder X-ray diffraction pattern of $Sm_2Ir_8Al_{31}$ and the corresponding Rietveld refinement. The inset shows single crystals on millimeter-grid paper.

Single-crystal X-ray diffraction (SCXRD) reveals that $Sm_2Ir_8Al_{31}$ crystallizes in the tetragonal space group $P\bar{4}2_1m$, with lattice parameters $a$ = 12.9169 Å and $c$ = 15.5739 Å, as shown in **Tables 1 and 2** and **Fig. 1*a***. The crystal structure consists of a three-dimensional framework of Ir and Al atoms surrounding the Sm atoms. Refinement of the Al site occupancies yields an actual composition of $Sm_2Ir_8Al_{30.8}$, slightly deviating from the ideal stoichiometry of $Sm_2Ir_8Al_{31}$. Energy-dispersive X-ray spectroscopy (EDS) was performed to independently determine the elemental composition. The EDS results (**Fig. S1**) agree well with the SCXRD refinement, confirming the nearly stoichiometric Sm:Ir:Al ratio of 2:8:31.

To further characterize the bulk sample and assess its phase purity, powder X-ray diffraction (PXRD) measurements were performed. The experimental diffraction pattern was analyzed by Rietveld refinement using the tetragonal $P\bar{4}2_1m$ structural model obtained from SCXRD, as shown in **Fig. 1*b***. The calculated diffraction pattern reproduces the major experimental reflections, with a weighted profile R-factor ($R_{wp}$) of 8.82%. No additional crystalline phases were detected within the sensitivity of the PXRD measurement, indicating that $Sm_2Ir_8Al_{31}$ is the dominant crystalline phase in the bulk sample. The inset of **Fig. 1*b*** shows representative single crystals of $Sm_2Ir_8Al_{31}$ with metallic color and well-defined crystal facets.

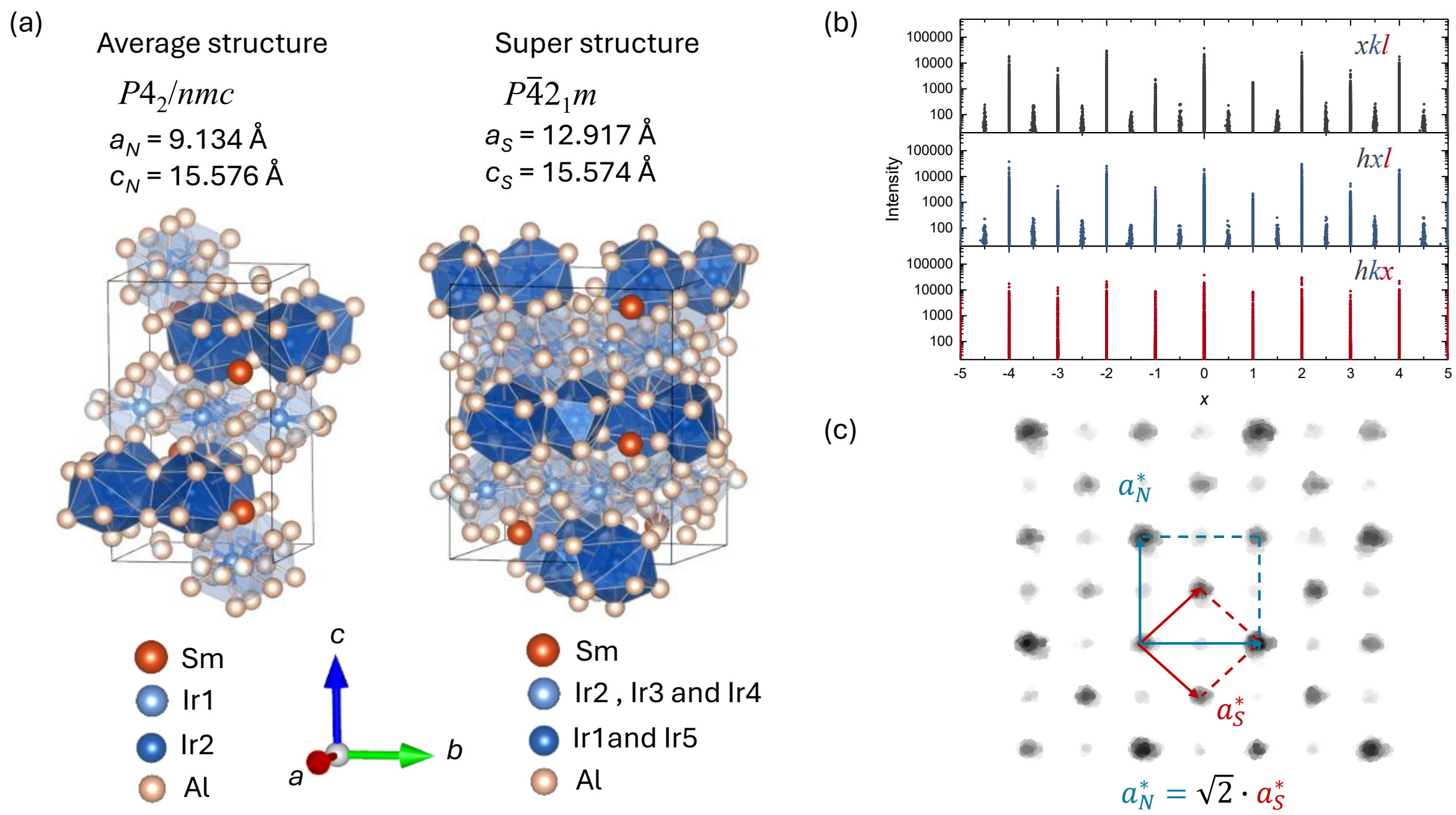

**Fig. 2| Crystalline structure of $Sm_2Ir_8Al_{31}$.** (***a***) Crystal structures shown using the average cell and the $\sqrt{2} \times \sqrt{2} \times 1$ supercell. (***b***) Reciprocal-space intensity of $Sm_2Ir_8Al_{31}$ along the *xkl*, *hxl*, and *hkx* directions. (***c***) Reciprocal-space intensity map illustrating the relationship between the average-cell and supercell reciprocal lattices. The average structure adopts the same structural framework as $UIr_4Al_{15}$, whereas the superstructure exhibits a $(\sqrt{2} \times \sqrt{2} \times 1)$ enlargement of the unit cell.

We examined the crystal structure of $Sm_2Ir_8Al_{31}$ using single-crystal X-ray diffraction. The strongest diffraction reflections can be indexed using a tetragonal unit cell with lattice parameters $a$ = 9.134 Å and $c$ = 15.576 Å, corresponding to the space group $P4_2/nmc$ shown in **Tables 1** and **2**. This average structure closely resembles that of $UIr_4Al_{15}$, sharing a similar arrangement of the rare-earth/actinide atoms within the Ir-Al framework **(Fig. 2*a*)**. However, careful examination of the diffraction data reveals additional reflections with intensities approximately two orders of magnitude weaker than those of the main reflections (**Fig. 2*b***). These weak reflections cannot be accounted for by the average structural model, indicating the presence of a larger periodicity.

Indexing these additional reflections yields a $(\sqrt{2} \times \sqrt{2} \times 1)$ supercell with lattice parameters a = 12.917 Å and c = 15.574 Å. The supercell is rotated 45° in the *ab* plane relative to the average cell, as shown in the reciprocal-space intensity map in **Fig. 2*c***. The diffraction data can be described using the noncentrosymmetric tetragonal space group $P\overline{4}2_1m$. The resulting superstructure retains the main Sm–Ir–Al framework of the average structure, while several atomic sites become crystallographically distinct because of the enlarged unit cell and reduced symmetry (**Fig. 2*a***).

The reciprocal-space intensity distributions along the $h$, $k$, and $l$ directions further illustrate the relationship between the average structure and the superstructure (**Fig. 2b**). Strong reflections correspond to the average structural periodicity, whereas much weaker superlattice reflections reveal additional ordering within the enlarged unit cell. These observations establish that $Sm_2Ir_8Al_{31}$ adopts a commensurate ($\sqrt{2} \times \sqrt{2} \times 1$) superstructure derived from a $UIr_4Al_{15}$-like average structure.

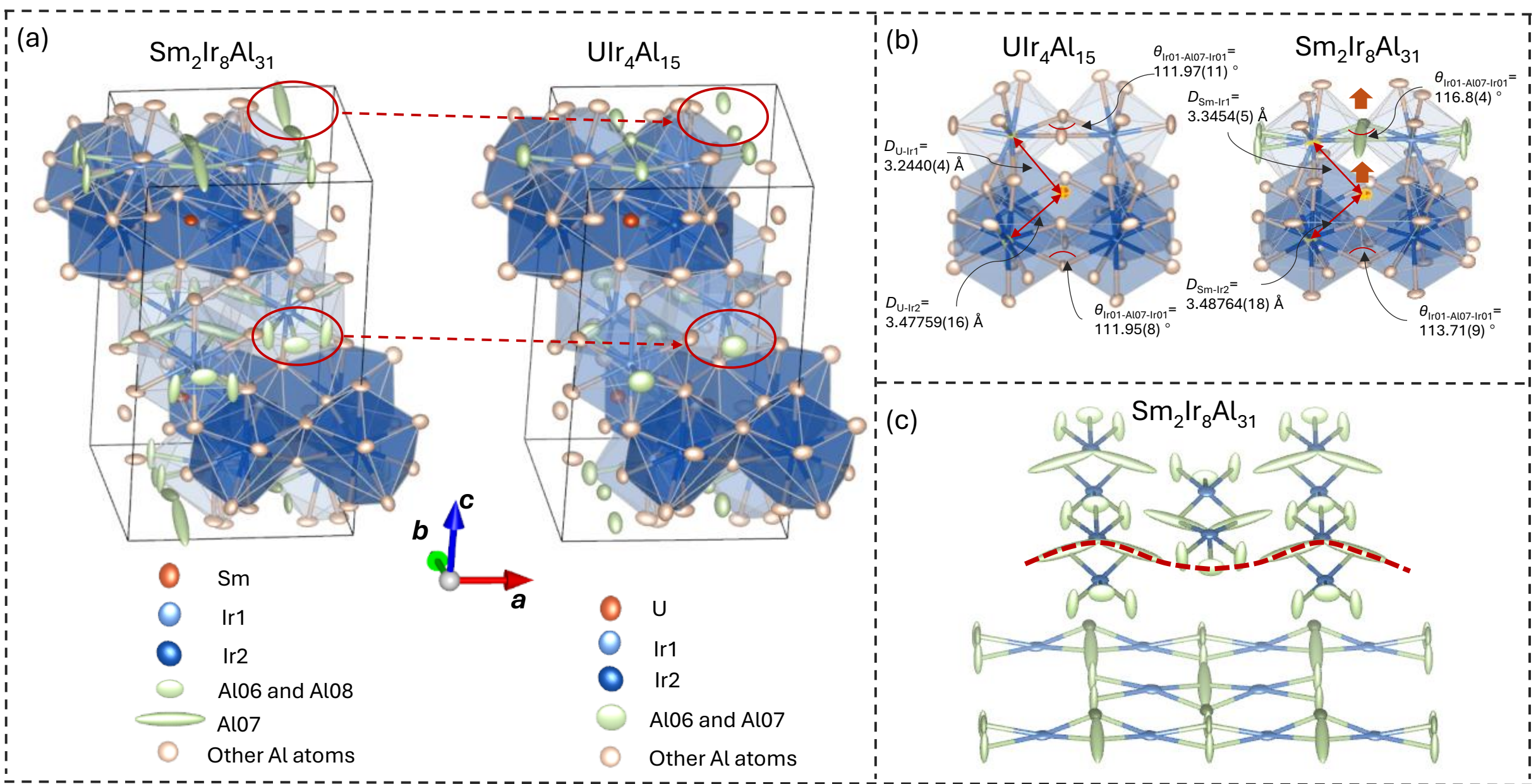

**Fig. 3 | The structure comparison between $Sm_2Ir_8Al_{31}$ and $UIr_4Al_{15}$.** (***a***) Comparison of the average structures, highlighting the Al sites with pronounced positional disorder. (***b***) Local structural distortion around the Sm/U and Ir framework, illustrated by selected interatomic distances and bond angles. (***c***) Anisotropic displacement and splitting of the Al sublattice in $Sm_2Ir_8Al_{31}$.

A direct comparison with $UIr_4Al_{15}$ reveals that replacement of U by Sm produces a substantial rearrangement of the local structure while preserving the overall Ir–Al framework in **Fig. 3*a***. The Sm compound has a more expanded lattice, which increases the available volume within the framework and modifies the local coordination geometry around the rare-earth/actinide site. However, this expansion is not accommodated uniformly by all Ir sites. The environment associated with Ir2 changes only modestly, whereas the coordination involving Ir1 shows a considerably larger rearrangement, as reflected by the changes in the Sm/U–Ir distances and the Ir–Al–Ir bond angles shown in **Fig. 3*b***. Thus, the structural response to Sm replacement is highly anisotropic rather than a simple uniform expansion of the parent framework.

This local distortion is most clearly manifested in the Al sublattice. In $UIr_4Al_{15}$, the corresponding Al06 and Al07 positions exhibit relatively regular thermal displacement parameters. In $Sm_2Ir_8Al_{31}$, the modified Ir1-centered coordination creates a less constrained local environment for these Al atoms. Consequently, the Al06 position becomes unstable and splits into two nearby positions, while Al07 develops a strongly elongated, cigar-shaped displacement ellipsoid in **Fig. 3*a*** and ***c***. These features indicate substantial positional disorder along a preferred direction and suggest that the expanded Sm-containing framework provides additional configurational space for Al. The partial occupation of the split Al sites therefore provides a natural structural origin for the higher Al content in $Sm_2Ir_8Al_{31}$ compared with stoichiometric $UIr_4Al_{15}$.

An electronic contribution may also be involved. The 4*f* electrons of Sm are substantially more localized than the 5*f* electrons of U, which generally participate more strongly in hybridization with the surrounding Ir–Al states. Replacing U with Sm therefore changes not only the size of the framework but also the electronic interaction between the *f*-element site and the Ir–Al network. The additional Al in $Sm_2Ir_8Al_{31}$ may consequently help stabilize an electron count and bonding configuration different from that of $UIr_4Al_{15}$. At present, however, the structural and electronic effects cannot be completely separated, and the observed Al enrichment likely reflects the combined influence of lattice expansion, local bonding rearrangement, and altered *f*-electron hybridization.

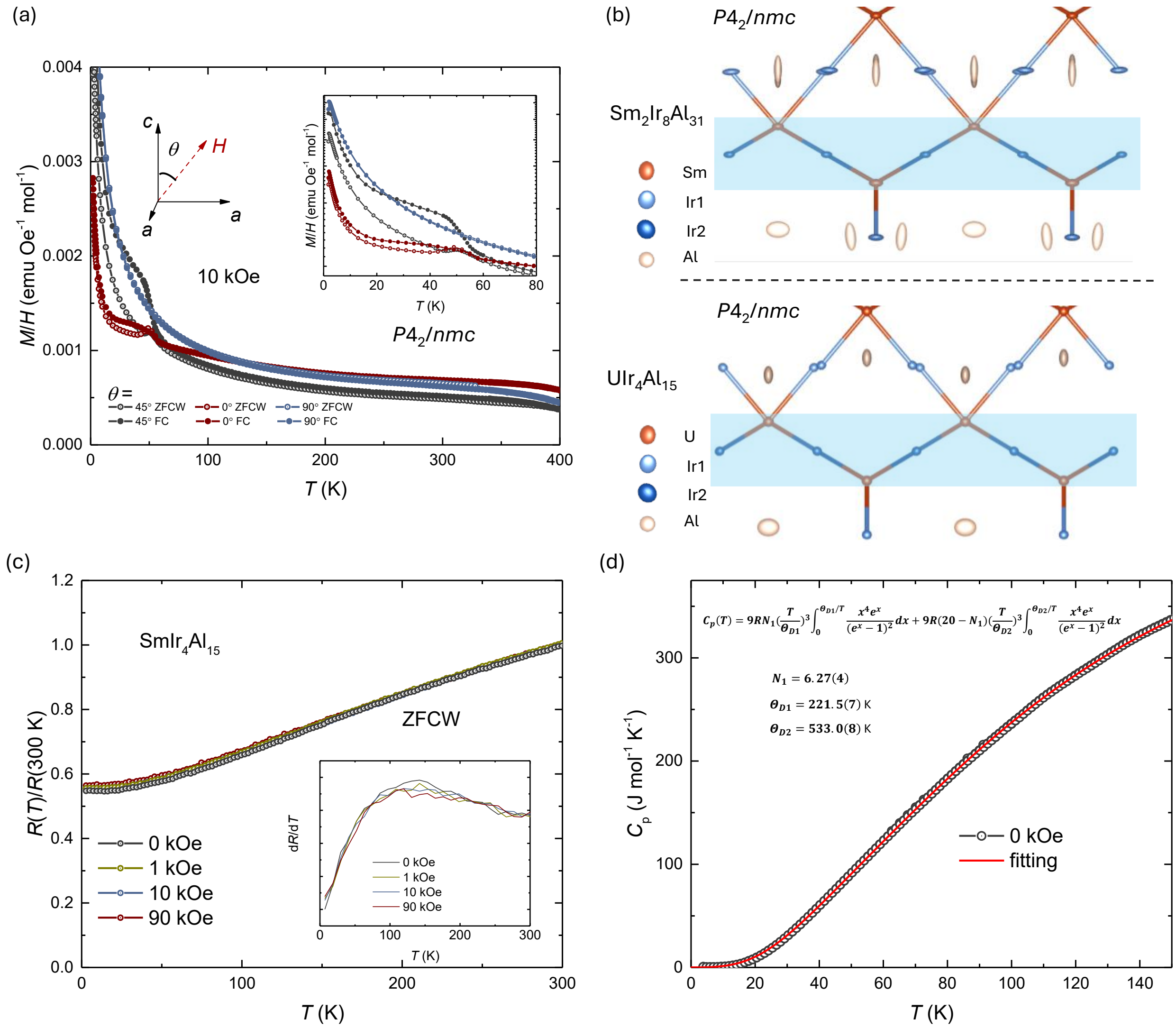

**Fig. 4 | Temperature-dependent properties of $Sm_2Ir_8Al_{31}$.** (***a***) magnetic susceptibility measured at 10 kOe for different field orientations under ZFCW and FC conditions; (***b***) Comparison of the atomic displacement ellipsoids of $Sm_2Ir_8Al_{31}$ (top) and $UIr_4Al_{15}$ (bottom), showing different anisotropic atomic displacement patterns; The blue shading highlights the in-plane atomic displacement of the Sm and Ir2 atoms within the *ab* plane, which show relatively small positional changes upon replacing U with Sm. (***c***) normalized electrical resistance under different magnetic fields, as shown in the inset; and (***d***) zero-field heat capacity, with the Debye mode fitting in the inset.

To compare with the structure of $UIr_4Al_{15}$, we use the average structural model with space group $P4_2/nmc$ for the temperature-dependent structural analysis and use its crystallographic axes as references for the anisotropic magnetization measurements. The temperature-dependent lattice parameters evolve smoothly between 100 and 330 K, with no detectable discontinuity or hysteresis between warming and cooling, as shown in **Fig. S2**. Electrical resistance also varies continuously with temperature over the measured range, and no sharp anomaly is resolved in either $R(T)$ or

d*R*/d*T, as* shown in **Fig. 4*c***. Similarly, in **Fig. 4*d***, the zero-field heat capacity is smooth and can be well described by a two-Debye model without an additional thermodynamic anomaly. These observations provide no evidence for a bulk structural or thermodynamic phase transition over this temperature range.

In contrast, the magnetic susceptibility of $Sm_2Ir_8Al_{31}$ exhibits a broad feature near 50 K that is most pronounced at $\theta = 45°$ and remains observable for $H \parallel c$ ($\theta = 0°$), whereas it is much less apparent at $\theta = 90°$, as shown in **Fig. 4*a***. Compared with the relatively sharp and strongly orientation-dependent magnetic anomalies observed in $UIr_4Al_{15}$, the feature in $Sm_2Ir_8Al_{31}$ occurs at a lower temperature and is considerably broader. Although the magnetic susceptibility remains anisotropic, the broad feature is observed for more than one field orientation, with its magnitude strongly dependent on the applied-field direction. Furthermore, the absence of a corresponding heat-capacity anomaly suggests that this feature is unlikely to originate from a conventional bulk magnetic phase transition involving a substantial entropy change.

To better understand the differences in magnetic behavior between these two compounds, we compared their atomic displacement ellipsoids using the same average structural model with space group $P4_2/nmc$, as shown in **Fig. 4*b***. In $UIr_4Al_{15}$, the displacement ellipsoids of U and Ir are nearly spherical, indicating relatively isotropic atomic displacement parameters. In contrast, Sm and several Ir sites in $Sm_2Ir_8Al_{31}$ show anisotropic displacement ellipsoids, predominantly elongated in the crystallographic *ab* plane, accompanied by changes in the local arrangement and displacement parameters of Al atoms. These differences suggest that the Sm-centered coordination environment exhibits a more anisotropic distribution of atomic displacements than the corresponding U-centered environment in $UIr_4Al_{15}$. Such anisotropic displacement parameters may reflect local structural distortions, anisotropic thermal motion, or positional disorder associated with the superstructure.

These structural differences may contribute to the distinct magnetic responses of the two compounds. In particular, changes in the local coordination environment can modify the crystal electric field (CEF) experienced by the $Sm^{3+}$ ions, thereby influencing magnetic anisotropy and temperature dependence of susceptibility. The broad feature near 50 K may be associated with the thermal population of low-lying CEF levels, although this possibility has not been experimentally established. Alternatively, short-range magnetic correlations may contribute to this behavior.

Changes in the local coordination environment may also influence magnetic interactions and contribute to the broad temperature dependence of the susceptibility. Although the present results do not establish a direct causal relationship between atomic displacements and magnetism, the comparison with $UIr_4Al_{15}$ suggests that differences in local structural environments may play an important role in determining their magnetic properties.

## Conclusion

$Sm_2Ir_8Al_{31}$ retains the basic tetragonal Ir–Al framework of $UIr_4Al_{15}$ but exhibits a $\sqrt{2} \times \sqrt{2} \times 1$ superstructure with reduced crystallographic symmetry. Compared with $UIr_4Al_{15}$, $Sm_2Ir_8Al_{31}$ shows pronounced anisotropic atomic displacement ellipsoids and changes in the local coordination environments. The Ir2-related framework remains relatively unchanged, whereas the Ir1-related network undergoes a more substantial rearrangement. These structural differences are accompanied by changes in the Al sublattice, including site splitting, elongated displacement ellipsoids, and partial occupancies, resulting in a refined composition of $Sm_2Ir_8Al_{30.8}$. The superstructure thus reflects a local rearrangement and ordering of the Al sublattice while largely preserving the underlying Ir–Al framework.

The magnetic properties of $Sm_2Ir_8Al_{31}$ also differ significantly from those of $UIr_4Al_{15}$. Although the magnetic susceptibility remains anisotropic, it shows a broad feature near 50 K across multiple field orientations, unlike the sharper, more strongly orientation-dependent anomaly in $UIr_4Al_{15}$ near 310 K. The absence of corresponding anomalies in heat capacity, electrical resistance, and lattice parameters suggests that the 50 K feature is more consistent with a magnetic crossover than a conventional bulk phase transition. This behavior may arise from the crystal-electric-field environment of $Sm^{3+}$, potentially accompanied by short-range magnetic correlations. The structural differences between the two compounds, together with the more localized nature of Sm 4*f* electrons compared with U 5*f* electrons, may contribute to their distinct magnetic responses. Overall, $Sm_2Ir_8Al_{31}$ demonstrates how modifications of the *f*-element site and local atomic arrangements can produce a superstructure and substantially alter magnetic behavior within a related Ir–Al structural framework.

## Acknowledgments

M.X. at Weiwei Xie's lab was supported by the U.S. Department of Energy (DOE), Division of Basic Energy Sciences under Contract DE-SC0023648.

## References

(1) Moore, K. T.; van der Laan, G. Nature of the 5f States in Actinide Metals. *Rev. Mod. Phys.* **2009**, *81* (1), 235–298. https://doi.org/10.1103/RevModPhys.81.235

(2) Lander, G. H.; Brooks, M. S. S.; Johansson, B. Orbital Band Magnetism in Actinide Intermetallics. *Phys. Rev. B* **1991**, *43* (16), 13672–13675. https://doi.org/10.1103/PhysRevB.43.13672

(3) Freeman, A. J.; Koelling, D. D. Electronic Structure and Properties of the Actinides. *Physica B+C* **1977**, *86–88*, 16–25. https://doi.org/10.1016/0378-4363(77)90211-X

(4) Zwicknagl, G.; Reese, M. Dual Nature of Strongly Correlated 5f Electrons. *J. Magn. Magn. Mater.* **2007**, *310* (2), 201–206. https://doi.org/10.1016/j.jmmm.2006.11.195

(5) Aléonard, R.; Morin, P. Quadrupolar Interactions in Rare Earth Intermetallics. *J. Magn. Magn. Mater.* **1990**, *84* (3), 255–263. https://doi.org/10.1016/0304-8853(90)90103-W

(6) Xu, M.; Lee, Y.; Ke, X.; Kang, M.-C.; Boswell, M.; Bud'ko, S. L.; Zhou, L.; Ke, L.; Li, M.; Canfield, P. C.; Xie, W. Giant Uniaxial Magnetocrystalline Anisotropy in $SmCrGe_3$. *J. Am. Chem. Soc.* **2024**, *146* (44), 30294–30302. https://doi.org/10.1021/jacs.4c10056

(7) Jia, S.; Bud'ko, S. L.; Samolyuk, G. D.; Canfield, P. C. Nearly Ferromagnetic Fermi-Liquid Behaviour in $YFe_2Zn_{20}$ and High-Temperature Ferromagnetism of $GdFe_2Zn_{20}$. *Nat. Phys.* **2007**, *3* (5), 334–338. https://doi.org/10.1038/nphys568

(8) Torikachvili, M. S.; Jia, S.; Mun, E. D.; Hannahs, S. T.; Black, R. C.; Neils, W. K.; Martien, D.; Bud'ko, S. L.; Canfield, P. C. Six Closely Related $YbT_2Zn_{20}$ (T = Fe, Co, Ru, Rh, Os, Ir) Heavy Fermion Compounds with Large Local Moment Degeneracy. *Proceedings of the National Academy of Sciences* **2007**, *104* (24), 9960–9963. https://doi.org/10.1073/pnas.0702757104

(9) Jia, S.; Ni, N.; Samolyuk, G. D.; Safa-Sefat, A.; Dennis, K.; Ko, H.; Miller, G. J.; Bud'ko, S. L.; Canfield, P. C. Variation of the Magnetic Ordering in $GdT_2Zn_{20}$ ( T=Fe, Ru, Os, Co, Rh and Ir) and Its Correlation with the Electronic Structure of Isostructural $YT_2Zn_{20}$. *Phys. Rev. B* **2008**, *77* (10), 104408. https://doi.org/10.1103/PhysRevB.77.104408

(10) Jia, S.; Ni, N.; Bud'ko, S. L.; Canfield, P. C. Magnetic Properties of $RFe_2Zn_{20}$ and $RCo_2Zn_{20}$ (R=Y, Nd, Sm, Gd, Lu). *Phys. Rev. B* **2009**, *80* (10), 104403. https://doi.org/10.1103/PhysRevB.80.104403

(11) Kangas, M. J.; Schmitt, D. C.; Sakai, A.; Nakatsuji, S.; Chan, J. Y. Structure and Physical Properties of Single Crystal $PrCr_2Al_{20}$ and $CeM_2Al_{20}$ (M=V, Cr): A Comparison of Compounds Adopting the $CeCr_2Al_{20}$ Structure Type. *J. Solid State Chem.* **2012**, *196*, 274–281. https://doi.org/10.1016/j.jssc.2012.06.035

(12) Thiede, V. M. T.; Jeitschko, W.; Niemann, S.; Ebel, T. $EuTa_2Al_{20}$, $Ca_6W_4Al_{43}$ and Other Compounds with $CeCr_2Al_{20}$ and $Ho_6Mo_4Al_{43}$ Type Structures and Some Magnetic Properties of These Compounds. *J. Alloys Compd.* **1998**, *267* (1–2), 23–31. https://doi.org/10.1016/S0925-8388(97)00532-X

(13) Moze, O.; Tung, L. D.; Franse, J. J. M.; Buschow, K. H. J. Crystal Structure and Magnetic Properties of $CeV_2Al_{20}$ and $CeCr_2Al_{20}$. *J. Alloys Compd.* **1998**, *268* (1–2), 39–41. https://doi.org/10.1016/S0925-8388(97)00586-0

(14) Niemann, S.; Jeitschko, W. Ternary Aluminides $AT_2Al_{20}$ (A = Rare Earth Elements and Uranium; T = Ti, Nb, Ta, Mo, and W) with $CeCr_2Al_{20}$-Type Structure. *J. Solid State Chem.* **1995**, *114* (2), 337–341. https://doi.org/10.1006/jssc.1995.1052

(15) Winiarski, M. J.; Klimczuk, T. Crystal Structure and Low-Energy Einstein Mode in $ErV_2Al_{20}$ Intermetallic Cage Compound. *J. Solid State Chem.* **2017**, *245*, 10–16. https://doi.org/10.1016/j.jssc.2016.09.029

(16) Kim, S. K.; Torikachvili, M. S.; Bud'ko, S. L.; Canfield, P. C. Search for Pressure-Induced Quantum Criticality in $YbFe_2Al_{20}$. *Phys. Rev. B* **2013**, *88* (4), 045116. https://doi.org/10.1103/PhysRevB.88.045116

(17) Kaluarachchi, U. S.; Xiang, L.; Ying, J.; Kong, T.; Struzhkin, V.; Gavriliuk, A.; Bud'ko, S. L.; Canfield, P. C. Collapse of the Kondo State and Ferromagnetic Quantum Phase Transition in $YbFe_2Al_{20}$. *Phys. Rev. B* **2018**, *98* (17), 174405. https://doi.org/10.1103/PhysRevB.98.174405

(18) Wörl, A.; Garst, M.; Yamane, Y.; Bachus, S.; Onimaru, T.; Gegenwart, P. Divergent Thermal Expansion and Grüneisen Ratio in a Quadrupolar Kondo Metal. *Phys. Rev. Res.* **2022**, *4* (2), L022053. https://doi.org/10.1103/PhysRevResearch.4.L022053

(19) Onimaru, T.; Matsumoto, K. T.; Inoue, Y. F.; Umeo, K.; Sakakibara, T.; Karaki, Y.; Kubota, M.; Takabatake, T. Antiferroquadrupolar Ordering in a Pr-Based Superconductor $PrIr_2Al_{20}$. *Phys. Rev. Lett.* **2011**, *106* (17), 177001. https://doi.org/10.1103/PhysRevLett.106.177001

(20) Isomae, T.; Sakai, A.; Fu, M.; Taniguchi, T.; Takigawa, M.; Nakatsuji, S. Extremely Large Magnetoresistance and Anisotropic Transport in the Multipolar Kondo System $PrTi_2Al_{20}$. *Phys. Rev. Res.* **2024**, *6* (1), 013009. https://doi.org/10.1103/PhysRevResearch.6.013009

(21) Lenk, M.; Gao, F.; Kroha, J.; Nevidomskyy, A. H. Strange-Metal Behavior without Fine Tuning in $PrV_2Al_{20}$. *Phys. Rev. Res.* **2024**, *6* (4), L042008. https://doi.org/10.1103/PhysRevResearch.6.L042008

(22) Patri, A. S.; Sakai, A.; Lee, S.; Paramekanti, A.; Nakatsuji, S.; Kim, Y. B. Unveiling Hidden Multipolar Orders with Magnetostriction. *Nat. Commun.* **2019**, *10* (1), 4092. https://doi.org/10.1038/s41467-019-11913-3

(23) Freyer, F.; Attig, J.; Lee, S.; Paramekanti, A.; Trebst, S.; Kim, Y. B. Two-Stage Multipolar Ordering in $PrT_2Al_{20}$ Kondo Materials. *Phys. Rev. B* **2018**, *97* (11), 115111. https://doi.org/10.1103/PhysRevB.97.115111

(24) Onimaru, T.; Kusunose, H. Exotic Quadrupolar Phenomena in Non-Kramers Doublet Systems — The Cases of $PrT_2Al_{20}$ ( T = Ir, Rh) and $PrT_2Al_{20}$ ( T = V, Ti). *J. Physical Soc. Japan* **2016**, *85* (8), 082002. https://doi.org/10.7566/JPSJ.85.082002

(25) Shimura, Y.; Tsujimoto, M.; Zeng, B.; Balicas, L.; Sakai, A.; Nakatsuji, S. Field-Induced Quadrupolar Quantum Criticality in $PrV_2Al_{20}$. *Phys. Rev. B* **2015**, *91* (24), 241102. https://doi.org/10.1103/PhysRevB.91.241102

(26) Tsujimoto, M.; Matsumoto, Y.; Tomita, T.; Sakai, A.; Nakatsuji, S. Heavy-Fermion Superconductivity in the Quadrupole Ordered State of $PrV_2Al_{20}$. *Phys. Rev. Lett.* **2014**, *113* (26), 267001. https://doi.org/10.1103/PhysRevLett.113.267001

(27) Matsubayashi, K.; Tanaka, T.; Sakai, A.; Nakatsuji, S.; Kubo, Y.; Uwatoko, Y. Pressure-Induced Heavy Fermion Superconductivity in the Nonmagnetic Quadrupolar System $PrTi_2Al_{20}$. *Phys. Rev. Lett.* **2012**, *109* (18), 187004. https://doi.org/10.1103/PhysRevLett.109.187004

(28) Sakai, A.; Kuga, K.; Nakatsuji, S. Superconductivity in the Ferroquadrupolar State in the Quadrupolar Kondo Lattice $PrTi_2Al_{20}$. *J. Physical Soc. Japan* **2012**, *81* (8), 083702. https://doi.org/10.1143/JPSJ.81.083702

(29) Sakai, A.; Nakatsuji, S. Kondo Effects and Multipolar Order in the Cubic $PrTr_2Al_{20}$ ( Tr =Ti, V). *J. Physical Soc. Japan* **2011**, *80* (6), 063701. https://doi.org/10.1143/JPSJ.80.063701

(30) Xu, M.; Klimczuk, T.; Maple, M. B.; Xie, W. Room-Temperature Magnetoelastic Coupling in $UIr_4Al_{15}$. **2026**.

(31) Tobash, P. H.; Scott, B. L.; Sidorov, V. A.; Ronning, F.; Gofryk, K.; Thompson, J. D.; Albers, R. C.; Zhu, J.-X.; Jones, M. D.; Bauer, E. D. Heavy Fermion Behavior in the New Antiferromagnetic Compound $UIr_4Al_{15}$. *J. Phys. Conf. Ser.* **2011**, *273*, 012061. https://doi.org/10.1088/1742-6596/273/1/012061

(32) de Wijn, H. W.; van Diepen, A. M.; Buschow, K. H. J. Nuclear Quadrupole Coupling and Crystal-Field Effects in the Cubic Intermetallic Compounds $ErAl_3$, Tm $Al_3$, and $YbAl_3$. *Phys. Rev. B* **1970**, *1* (11), 4203–4208. https://doi.org/10.1103/PhysRevB.1.4203

(33) Shimura, Y.; Zhang, Q.; Zeng, B.; Rhodes, D.; Schönemann, R.; Tsujimoto, M.; Matsumoto, Y.; Sakai, A.; Sakakibara, T.; Araki, K.; Zheng, W.; Zhou, Q.; Balicas, L.; Nakatsuji, S. Giant Anisotropic Magnetoresistance Due to Purely Orbital Rearrangement in the Quadrupolar Heavy Fermion Superconductor $PrV_2Al_{20}$. *Phys. Rev. Lett.* **2019**, *122* (25), 256601. https://doi.org/10.1103/PhysRevLett.122.256601

(34) Freyer, F.; Lee, S.; Kim, Y. B.; Trebst, S.; Paramekanti, A. Thermal and Field-Induced Transitions in Ferroquadrupolar Kondo Systems. *Phys. Rev. Res.* **2020**, *2* (3), 033176. https://doi.org/10.1103/PhysRevResearch.2.033176

(35) Canfield, P. C. NEW MATERIALS PHYSICS Paul C. Canfield Ames Laboratory and Department of Physics Iowa State University, Ames, Iowa 50011. No. 1.

(36) LSP INDUSTRIAL CERAMICS. Canfield Crucible Sets. **2015**.

(37) Parkin, S.; Moezzi, B.; Hope, H. XABS 2: An Empirical Absorption Correction Program. *J. Appl. Crystallogr.* **1995**, *28* (1), 53–56. https://doi.org/10.1107/S0021889894009428

(38) Walker, N.; Stuart, D. An Empirical Method for Correcting Diffractometer Data for Absorption Effects. *Acta Crystallogr. A* **1983**, *39* (1), 158–166. https://doi.org/10.1107/S0108767383000252

(39) Sheldrick, G. M. SHELXT – Integrated Space-Group and Crystal-Structure Determination. *Acta Crystallogr. A Found. Adv.* **2015**, *71* (1), 3–8. https://doi.org/10.1107/S2053273314026370

(40) Sheldrick, G. M. Crystal Structure Refinement with SHELXL. *Acta Crystallogr. C Struct. Chem.* **2015**, *71* (1), 3–8. https://doi.org/10.1107/S2053229614024218

(41) Dolomanov, O. V.; Bourhis, L. J.; Gildea, R. J.; Howard, J. A. K.; Puschmann, H. OLEX2 : A Complete Structure Solution, Refinement and Analysis Program. *J. Appl. Crystallogr.* **2009**, *42* (2), 339–341. https://doi.org/10.1107/S0021889808042726

(42) Toby, B. H.; Von Dreele, R. B. GSAS-II : The Genesis of a Modern Open-Source All Purpose Crystallography Software Package. *J. Appl. Crystallogr.* **2013**, *46* (2), 544–549. https://doi.org/10.1107/S0021889813003531

## Support Information

# Anisotropic Structural and Magnetic Response in $Sm_2Ir_8Al_{31}$: Al-Sublattice Reconstruction upon Sm Substitution

Mingyu Xu[1], Weiwei Xie[1*]

1. Department of Chemistry, Michigan State University, East Lansing, MI, 48824 USA

*Corresponding author: Dr. Weiwei Xie (xieweiwe@msu.edu)

## Table of Contents

**Table S1.** Crystal structure and refinement data for $Sm_2Ir_8Al_{31}$ at 300 K and refinement of the supercell. Values in parentheses represent estimated standard deviations from the refinements.

| | Normal cell | Supercell |
|---|---|---|
| | $Sm_2Ir_8Al_{31}$ | $Sm_2Ir_8Al_{31}$ |
| **Temperature (K)** | 300(2) | 300(2) |
| **Space Group** | $P4_2/nmc$ | $P\bar{4}2_1m$ |
| **Unit cell dimensions** | $a$ = 9.13411(9) Å; $b$ = 9.13411(9) Å; $c$ = 15.5755(2) Å. | $a$ = 12.9169(1) Å; $b$ = 12.9169(1) Å; $c$ = 15.5739(2) Å. |
| **Volume (Å$^3$)** | 1299.49(3) | 2598.45(4) |
| **Z** | 2 | 4 |
| **Density (Calculated) (g/m$^3$)** | 6.819 | 6.829 |
| **Absorption coefficient (mm$^{-1}$)** | 46.080 | 46.411 |
| **F(000)** | 2270.2 | 4546.5 |
| **Reflections collected** | 34914 | 74164 |
| **Independent reflections** | 1642 | 6162 |
| **Data/Restraints/Parameters** | 1642/0/68 | 6162/0/221 |
| **Final R indices** | $R_1$ = 0.0289, $wR_2$ = 0.072 | $R_1$ = 0.0330, $wR_2$ = 0.0813 |
| **Largest diff. peak and hole (e Å$^{-3}$)** | 6.38/-4.01 | 3.50/-3.86 |
| **Goodness-of-fit on F$^2$** | 1.010 | 1.036 |

**Table S2.** Atomic coordinates for $Sm_2Ir_8Al_{31}$ at 300 K and refinement of the supercell.

| | Atom | Wyck. | *x* | *y* | *z* | Occ. | $U_{eq}$ |
|---|---|---|---|---|---|---|---|
| Normal cell | Sm | 4*d* | 3/4 | 3/4 | 0.66545(3) | 1 | 0.0068(1) |
| | Ir1 | 8*g* | 3/4 | 0.51106(4) | 0.50266(17) | 1 | 0.0099(1) |
| | Ir2 | 8*f* | 0.00153(2) | 0.50153(2) | 3/4 | 1 | 0.0051(1) |
| | Al01 | 8*g* | 3/4 | 0.4016(3) | 0.65703(16) | 1 | 0.010(1) |
| | Al02 | 8*g* | 1/4 | 0.5993(3) | 0.81273(15) | 1 | 0.0076(4) |
| | Al03 | 8*g* | 3/4 | 0.5614(3) | 0.33771(15) | 1 | 0.010(1) |
| | Al04 | 8*g* | 0.0135(3) | 1/4 | 0.68211(16) | 1 | 0.009(1) |
| | Al05 | 16*h* | 0.9966(2) | 0.5282(2) | 0.58478(11) | 1 | 0.011(1) |
| | Al06 | 4*c* | 3/4 | 1/4 | 0.4739(4) | 0.58(2) | 0.014(2) |
| | Al07 | 8*g* | 0.8913(9) | 3/4 | 0.4791(2) | 1 | 0.062(2) |
| | Al08 | 8g | 0.6206(7) | 1/4 | 0.5050(6) | 0.40(1) | 0.016(2) |
| Super cell | Sm1 | 4*e* | 0.2495(3) | 0.7495(3) | 0.5846(5) | 1 | 0.0057(1) |
| | Sm2 | 4*e* | 0.2471(3) | 0.7471(3) | 0.0845(5) | 1 | 0.0043(1) |
| | Ir1 | 8*f* | 0.2483(3) | 0.0001(6) | 0.5000(2) | 1 | 0.0054(1) |
| | Ir2 | 4*e* | 0.6355(3) | 0.1355(3) | 0.7477(3) | 1 | 0.0046(1) |
| | Ir3 | 4*e* | 0.1250(3) | 0.6250(3) | 0.7467(3) | 1 | 0.0061(1) |
| | Ir4 | 8*f* | 0.3654(3) | 0.6265(3) | 0.2475(2) | 1 | 0.0065(1) |
| | Ir5 | 8*f* | 0.0027(5) | 0.2514(3) | 0.0003(2) | 1 | 0.0026(1) |
| | Al01 | 4*e* | 0.1317(2) | 0.6317(2) | 0.4326(3) | 1 | 0.0054(8) |
| | Al02 | 8*f* | 0.0748(3) | 0.0748(3) | 0.4399(2) | 1 | 0.010(1) |
| | Al03 | 8*f* | 0.5216(2) | 0.2400(3) | 0.1629(2) | 1 | 0.0078(6) |
| | Al04 | 4*e* | 0.0755(2) | 0.5755(2) | 0.5877(3) | 1 | 0.0030(6) |
| | Al05 | 4*e* | 0.6575(3) | 0.1575(3) | 0.9112(3) | 1 | 0.0095(8) |
| | Al06 | 4*e* | 0.5742(2) | 0.0742(2) | 0.5958(3) | 1 | 0.0040(7) |
| | Al07 | 8*f* | 0.4201(3) | 0.5742(3) | 0.0946(3) | 1 | 0.0124(7) |
| | Al08 | 4*e* | 0.5766(2) | 0.0766(2) | 0.0649(2) | 1 | 0.0023(7) |
| | Al09 | 8*f* | 0.1582(2) | 0.1564(2) | 0.5859(2) | 1 | 0.0069(6) |
| | Al10 | 8*f* | 0.2373(3) | 0.0139(3) | 0.3323(2) | 1 | 0.0092(7) |
| | Al11 | 8*f* | 0.1337(3) | 0.1285(3) | 0.0705(2) | 1 | 0.0091(7) |
| | Al12 | 8*f* | 0.0164(3) | 0.2383(3) | 0.3324(2) | 1 | 0.0087(6) |
| | Al13 | 8*f* | 0.2345(3) | 0.5109(3) | 0.1625(2) | 1 | 0.0070(6) |
| | Al14 | 4*e* | 0.1501(2) | 0.6501(2) | 0.9094(3) | 1 | 0.0074(8) |
| | Al15 | 4*e* | 0.1620(2) | 0.6620(2) | 0.2638(2) | 1 | 0.0072(6) |
| | Al16 | 4*e* | 0.6334(2) | 0.1334(2) | 0.4369(3) | 1 | 0.0029(7) |
| | Al17 | 4*e* | 0.0754(2) | 0.5754(2) | 0.0654(2) | 1 | 0.0021(7) |
| | Al18 | 8*f* | 0.1820(4) | 0.1778(4) | 0.2285(3) | 1 | 0.044(2) |
| | Al19 | 4*d* | 0 | 0 | 0.2760(4) | 0.74(2) | 0.018(2) |
| | Al20 | 4*e* | 0.5647(2) | 0.0647(2) | 0.2561(4) | 1 | 0.020(1) |
| | Al21 | 4*e* | 0.6993(2) | 0.1993(2) | 0.2775(3) | 1 | 0.011(1) |
| | Al22 | 2*c* | 0 | 1/2 | 0.2242(4) | 1 | 0.013(1) |
| | Al23 | 8*f* | 0.0672(5) | 0.0616(6) | 0.2474(7) | 0.32(1) | 0.011(3) |

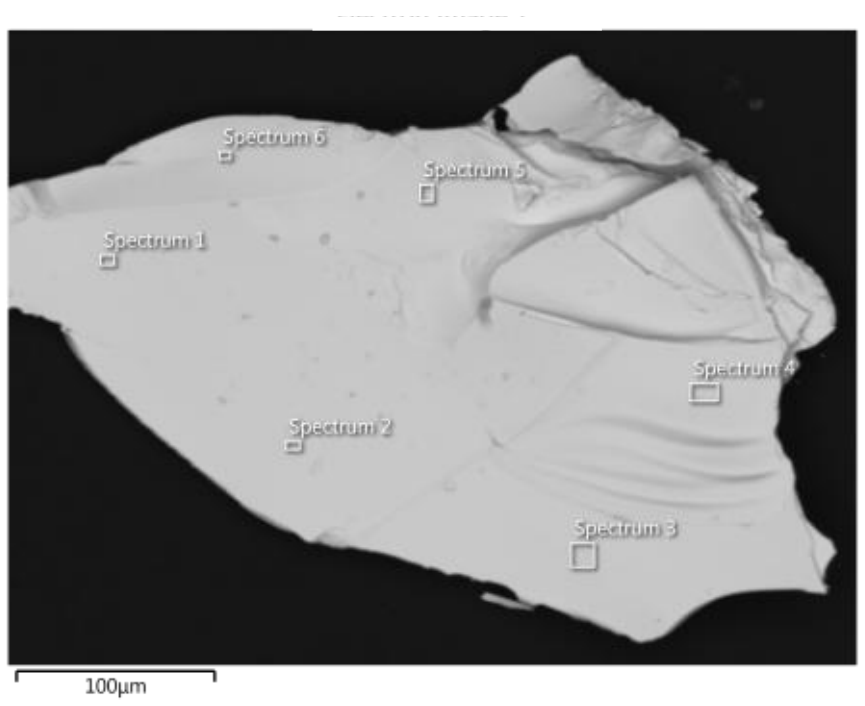

| | at% | normalized |
|---|---|---|
| Sm | 75.5 ± 1.0 | 30.8± 0.2 |
| Ir | 19.5 ± 0.8 | 8.0 ± 0.2 |
| Al | 4.9 ± 0.2 | 2.0 ± 0.04 |

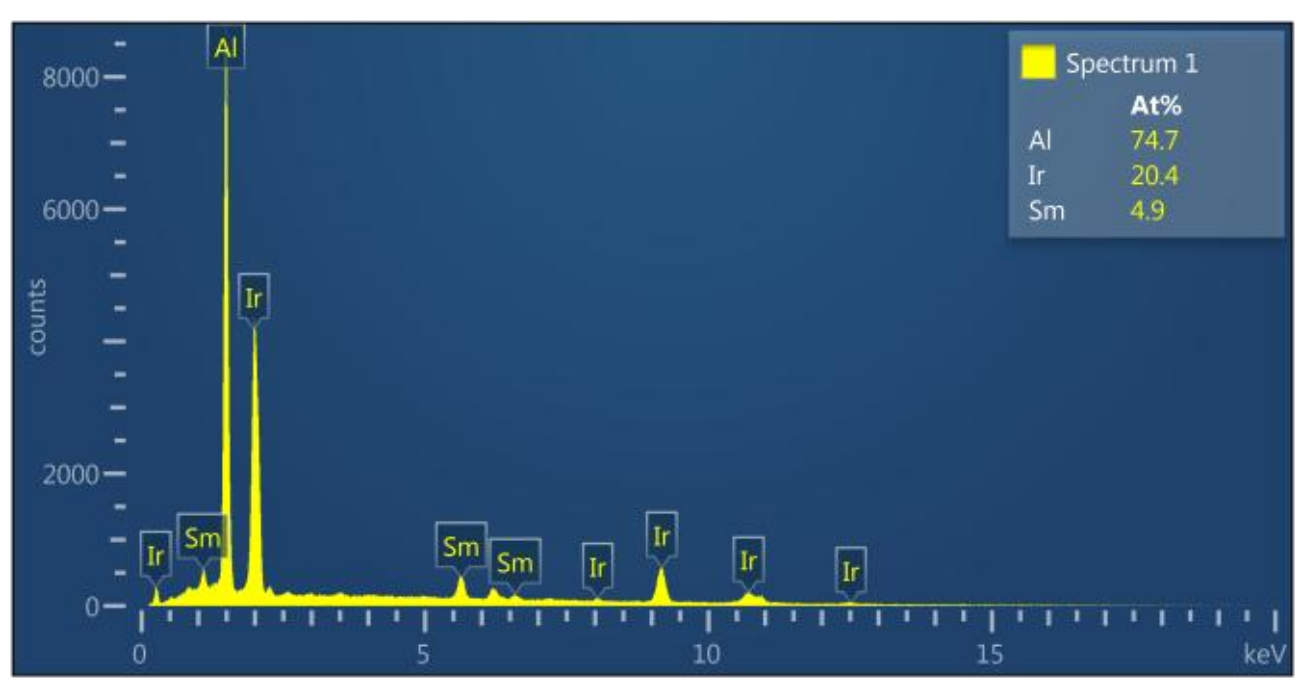

**Figure S1| Scanning Electron Microscope and Energy-Dispersive Spectroscopy of $Sm_2Ir_8Al_{31}$.**

**Scanning Electron Microscopy and Energy-Dispersive Spectroscopy of $Sm_2Ir_8Al_{31}$.**

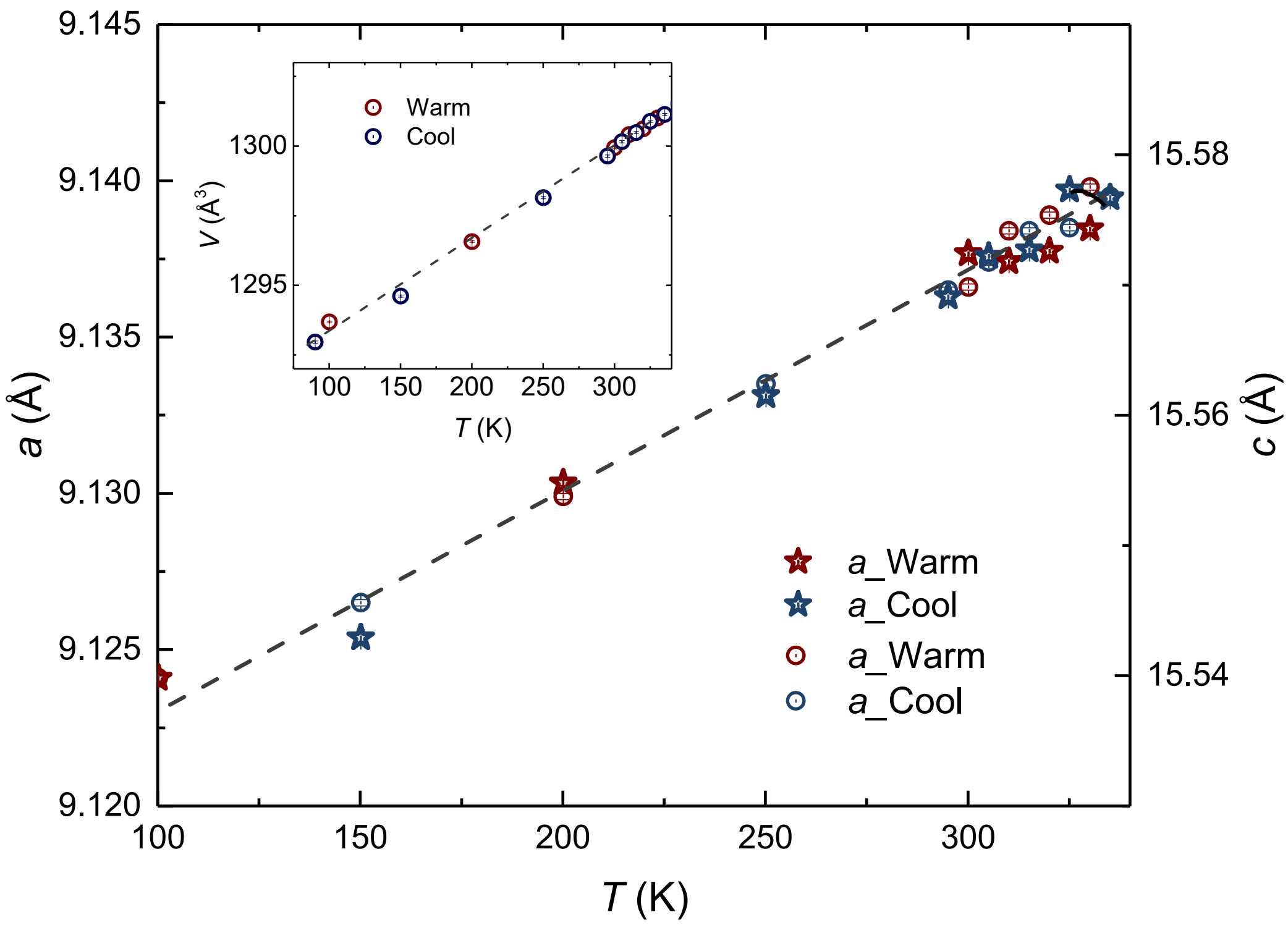

**Figure S2. Temperature-dependent lattice parameters of $Sm_2Ir_8Al_{31}$.**